\documentclass[10pt]{article}
\usepackage{amsmath}
\usepackage{amssymb}
\usepackage{enumerate}
\usepackage{epsf}
\usepackage{psfrag}
\usepackage{mathrsfs}
\usepackage{hyperref}
\usepackage{slashed}
\usepackage{mathrsfs}
\usepackage{amsthm} 
\newtheorem{innercustomthm}{Theorem}
\newenvironment{customthm}[1]
  {\renewcommand\theinnercustomthm{#1}\innercustomthm}
  {\endinnercustomthm}
\DeclareGraphicsExtensions{.eps,.art,.ART,.ps}
\newcommand{\bbR}{\mathbb{R}}      

\newcommand{\grad}{\operatorname{grad}}

\newtheorem{Thm}{Theorem}[section]

\theoremstyle{definition}

\newtheorem{Remark}[Thm]{Remark}

\begin{document}
\title{Decay of solutions of the wave equation in expanding cosmological spacetimes}
\author{Jo\~{a}o L.~Costa$^{1,2}$, Jos\'e Nat\'ario$^{2}$ and Pedro Oliveira$^{2}$\\ \\
{\small $^1$ Instituto Universit\'{a}rio de Lisboa (ISCTE-IUL), Lisboa, Portugal}\\
{\small $^2$ CAMGSD, Departamento de Matem\'{a}tica, Instituto Superior T\'{e}cnico,}\\
{\small Universidade de Lisboa, Portugal}
}
\date{}
\maketitle
\begin{abstract}
We study the decay of solutions of the wave equation in some expanding cosmological spacetimes, namely flat Friedmann-Lema\^\i tre-Robertson-Walker (FLRW) models and the cosmological region of the Reissner-Nordstr\"{o}m-de Sitter (RNdS) solution. By introducing a partial energy and using an iteration scheme, we find that, for initial data with finite higher order energies, the decay rate of the time derivative is faster than previously existing estimates. For models undergoing accelerated expansion, our decay rate appears to be (almost) sharp.
\end{abstract}
\tableofcontents
%
%
%
\section{Introduction and main results}\label{section0}
The purpose of this paper is to obtain the exact decay rates for solutions of the wave equation in expanding cosmological spacetimes. We are particularly interested in the decay rates of time derivatives, because they characterize the asymptotics of solutions in spacetimes with a spacelike null infinity $\mathscr{I}^+$. Moreover, if one regards the wave equation as a proxy for the Einstein equations, then time derivatives are akin to second fundamental forms of time slices, and their decay rates may be useful in formulating and proving cosmic no-hair theorems (in the spirit of \cite{AndreassonRingstrom}). In fact, the results of the present paper provided important insights for the recent analysis of the cosmic no-hair conjecture in spherically symmetric black hole spacetimes in \cite{CostaNatarioOliveira}.

A physical argument about which decay to expect can be made by considering an expanding FLRW model with flat $n$-dimensional spatial sections of radius $a(t)$. On one hand, the energy density of a solution $\phi$ of the wave equation is of the order of $(\partial_t \phi)^2$. On the other hand, if the wavelength of the particles associated with $\phi$ follows the expansion, then it is proportional to $a(t)$, and so their energy varies as $a(t)^{-1}$. Therefore, the energy density should behave like $a(t)^{-n-1}$, and we would expect $\partial_t\phi$ to decay as $a(t)^{-\frac{n+1}2}$. We shall see that, in reality, things are more complicated: this decay rate only seems to hold for spacetimes which are expanding sufficiently slowly, and, in particular, do not have a spacelike $\mathscr{I}^+$.

There are few results in the literature about this problem. Klainerman and Sarnak \cite{KlainermanSarnak} gave the explicit solution for the wave equation in FLRW models corresponding to dust matter, zero cosmological constant and $3$-dimensional flat or hyperbolic spatial sections. This was used in \cite{AbbasiCraig} to show that, in the flat case, solutions with initial data of compact support decay as $t^{-1}$, that is, as $a(t)^{-\frac32}$ (but without decay estimates for the time derivative). The same problem was studied further in \cite{GalstianKinoshitaYagdjian, GalstianYagdjian}, including $L^p$-$L^q$ decay estimates and paramatrices. The wave equation in the de Sitter spacetime with flat $3$-dimensional spatial sections was analyzed by Rendall \cite{Rendall}; he proved that the time derivative decays at least as $e^{-Ht}$ (with $H=\sqrt{\Lambda/3}$ being the Hubble constant, where $\Lambda>0$ is the cosmological constant), that is, as $a(t)^{-1}$, and conjectured a decay of order $e^{-2Ht}$, that is, $a(t)^{-2}$. This was also the decay found in \cite{CostaAlhoNatario} for spherical waves when approaching $i^+$. This problem was studied further in \cite{YagdjianGalstian}, including $L^p$-$L^q$ decay estimates. Recently, an extensive study of systems of linear wave equations on various cosmological backgrounds was presented in the monograph~\cite{Ringstrom}.

Our main result in the FLRW setting is the following:

\begin{customthm}{1} \label{mainthm1}
Consider an expanding FLRW model with flat $n$-dimensional spatial sections ($n\geq 2$), given by $I \times \bbR^n$ ($I \subset \bbR$ being an open an interval) with the metric
\begin{equation}
g = -dt^2 + a^2(t) \left(\left(dx^1\right)^2 + \ldots + \left(dx^n\right)^2 \right) \, ,
\end{equation}
where $a(t)$ satisfies $\dot{a}(t) \geq 0$ for $t \geq t_0$. Let $\phi$ be a smooth solution of the Cauchy problem
\begin{equation}
\begin{cases}
\Box_g \phi = 0 \\
\phi(t_0,x) = \phi_0(x) \\
\partial_t\phi(t_0,x) = \phi_1(x)
\end{cases}
\, ,
\end{equation}
and suppose that there exists $k>\frac{n}2 + 2$ such that
\begin{equation}
\| \phi_0 \|_{H^k(\bbR^n)} < + \infty \quad \text{ and } \quad \| \phi_1 \|_{H^{k-1}(\bbR^n)} < + \infty \, . \label{finiteSobolev1}
\end{equation}
Then, given $\delta > 0$, we have, for all $t \geq t_0$,
\begin{equation}
\|\partial_t\phi(t,\cdot)\|_{L^\infty(\bbR^n)} \lesssim a(t)^{-2+\varepsilon+\delta} \, , \label{mainestimate}
\end{equation}
where $\varepsilon>0$ is any positive number such that
\begin{equation}
\int_{t_0}^{+\infty} a(t)^{-\varepsilon} dt < +\infty \, .
\end{equation}
\end{customthm}

\begin{Remark}
For de Sitter's spacetime, in particular, we have $a(t) = e^{Ht}$ (with  $H=\sqrt{2\Lambda/(n(n-1))}$ being the Hubble constant, where $\Lambda>0$ is the cosmological constant), and so $\varepsilon$ can be chosen arbitrarily small. Therefore, for any $\delta>0$ we have
\begin{equation}
| \partial_t\phi | \lesssim a(t)^{-2+\delta} = e^{-(2-\delta)Ht} \, , \label{mainestimatedS}
\end{equation}
in agreement with Rendall's conjecture (up to the small quantity $\delta>0$). Note that this does not agree with the na{\"\i}ve physical expectation above, except, by coincidence, for $n=3$.  We show in Appendix~\ref{appendixA} that this decay rate is (almost) sharp.
\end{Remark}

\begin{Remark}
For $a(t)=t^p$ we have $\varepsilon = \frac{1+\delta}{p}$ for any $\delta > 0$, and so Theorem~\ref{mainthm1} gives
\begin{equation}
| \partial_t\phi | \lesssim a(t)^{-2+\frac{1+\delta}{p}} = t^{-(2p-1-\delta)} \, . \label{mainestimatep}
\end{equation}
Again, this does not conform to the na{\"\i}ve physical expectation.  We show in Appendix~\ref{appendixA} that this decay rate is (almost) sharp for $p>1$, that is, for the case where the expansion is accelerating, and there exists a spacelike $\mathscr{I}^+$. For $p<1$, that is, for the case where the expansion is decelerating, the decay rate given by Theorem~\ref{mainthm1} is poor, and in fact the exponent $-2+\frac{1+\delta}{p}$ can be easily improved to $-1$. However, the mode calculations in Appendix~\ref{appendixA} strongly suggest that in this case the relevant exponent should be the one coming from the na{\"\i}ve physical argument, namely $-\frac{n+1}2$.
\end{Remark}

\begin{Remark}
It will be clear from the proof that this result can easily be generalized to expanding FLRW models whose spatial sections have different geometries and/or topologies. In particular, it is true for compact spatial sections, showing that the decay mechanism is the cosmological expansion, and not dispersion.
\end{Remark}

A second important class of cosmological spacetimes is given by the Reissner-Nordstr\"{o}m-de Sitter solution (Schwarzschild-de Sitter being a particular case). The behavior of linear waves in the static region of these solutions has been studied (in the mathematics literature) in the work of Dafermos-Rodnianski \cite{DafermosRodnianski}, Bony-H\"{a}fner \cite{BonyHafner}, Melrose-S\'{a} Baretto-Vasy \cite{MelroseBarretoVasy} and Dyatlov \cite{Dyatlov, Dyatlov2}. Schlue~\cite{Schlue} studied the wave equation in the cosmological region of the Schwarzschild-de Sitter (also Kerr-de Sitter) solution. He obtained a decay of at least $\frac{1}{r^2}$ for the $r$ derivative\footnote{Note that $r$ is a time coordinate in the cosmological region.} of solutions of the wave equation as $r \to +\infty$.

Our main result in this setting reads as follows:

\begin{customthm}{2} \label{mainthm2}
Consider an $(n+1)$-dimensional sub-extremal Reissner-Nordstr\"{o}m-de Sitter solution ($n \geq 3$), given by the metric
\begin{equation}
g = - \left(r^2 + \frac{2M}{r^{n-2}} - \frac{e^2}{r^{n-1}} - 1\right)^{-1} dr^2 + \left(r^2 + \frac{2M}{r^{n-2}} - \frac{e^2}{r^{n-1}} - 1\right) dt^2 + r^2 d\Omega^2 \, ,
\end{equation}
where $d\Omega^2$ represents the metric of the unit $(n-1)$-dimensional sphere $S^{n-1}$, the constants $M$ and $e$ are proportional to the mass and charge of the black holes, and we have set the cosmological constant equal to $\frac12 n (n-1)$ by an appropriate choice of units. Let $\phi$ be a smooth solution of the wave equation
\begin{equation}
\Box_g \phi = 0 \, ,
\end{equation}
and suppose that there exists $k>\frac{n}2+2$ such that
\begin{equation} \label{hypo}
\| \phi \|_{H^k(\mathcal{CH}_1^+)} < + \infty \quad \text{ and } \quad \| \phi \|_{H^k(\mathcal{CH}_2^+)} < + \infty \, ,
\end{equation}
where $\mathcal{CH}_1^+ \cong \mathcal{CH}_2^+ \cong \bbR \times S^{n-1}$ are the two connected components of the future cosmological horizon, of radius $r_c$, parameterized by the flow parameter of the global Killing vector field $\frac{\partial}{\partial t}$. Then, given $\delta > 0$, we have, for all $r \geq r_0 > r_c$,
\begin{equation}
\|\partial_r\phi(r,\cdot)\|_{L^\infty(\bbR\times S^{n-1})} \lesssim {r}^{-3+\delta} \, .
\end{equation}
\end{customthm}

\begin{Remark}
This is the decay rate one would expect from Rendall's conjecture, since for free-falling observers in the cosmological region one has $r(\tau) \sim e^\tau \sim a(\tau)$, where $\tau$ is the proper time and $a(\tau)$ the radius of a comparable FLRW model, so that $\partial_r \phi \sim \partial_\tau \phi / \partial_\tau r \sim e^{-2\tau} / e^{\tau}$. The hypotheses in \eqref{hypo} can be recovered from the (higher dimensional version of the) analysis of the static region in \cite{DafermosRodnianski}.
\end{Remark}
%
%
\section{Decay in FLRW: Proof of Theorem~\ref{mainthm1}}\label{section1}
In this section we present the proof of Theorem~\ref{mainthm1}. For the reader's convenience we break it up into elementary steps.
%
%
\subsection{Wave equation in FLRW}\label{subsection1.1}
Consider an expanding FLRW model with flat $n$-dimensional spatial sections, given by the metric
\begin{equation}
g = -dt^2 + a^2(t) \left(\left(dx^1\right)^2 + \ldots + \left(dx^n\right)^2 \right) \, ,
\end{equation}
with $\dot{a}(t) \geq 0$ for $t \geq t_0$. The wave equation in this background,
\begin{equation}
\Box_g \phi = 0 \Leftrightarrow \partial_\mu \left(\sqrt{-g} \, \partial^\mu \phi \right) = 0 \Leftrightarrow \partial_\mu \left(a^n \partial^\mu \phi \right) = 0 \, ,
\end{equation}
can be written as
\begin{equation} \label{waveqn}
- \ddot{\phi} - \frac{n\dot{a}}{a} \dot{\phi} + \frac1{a^2} \delta^{ij} \partial_i \partial_j \phi = 0 \, ,
\end{equation}
where the dot denotes differentiation with respect to $t$ and the latin indices $i$ and $j$ run from $1$ to $n$.
%
%
\subsection{Energy}\label{subsection1.2}
\begin{figure}[h!]
\begin{center}
\psfrag{t=t0}{$t=t_0$}
\psfrag{t=t1}{$t=t_1$}
\psfrag{B}{$\mathcal{B}$}
\psfrag{R}{$\mathcal{R}$}
\epsfxsize=1.0\textwidth
\leavevmode
\epsfbox{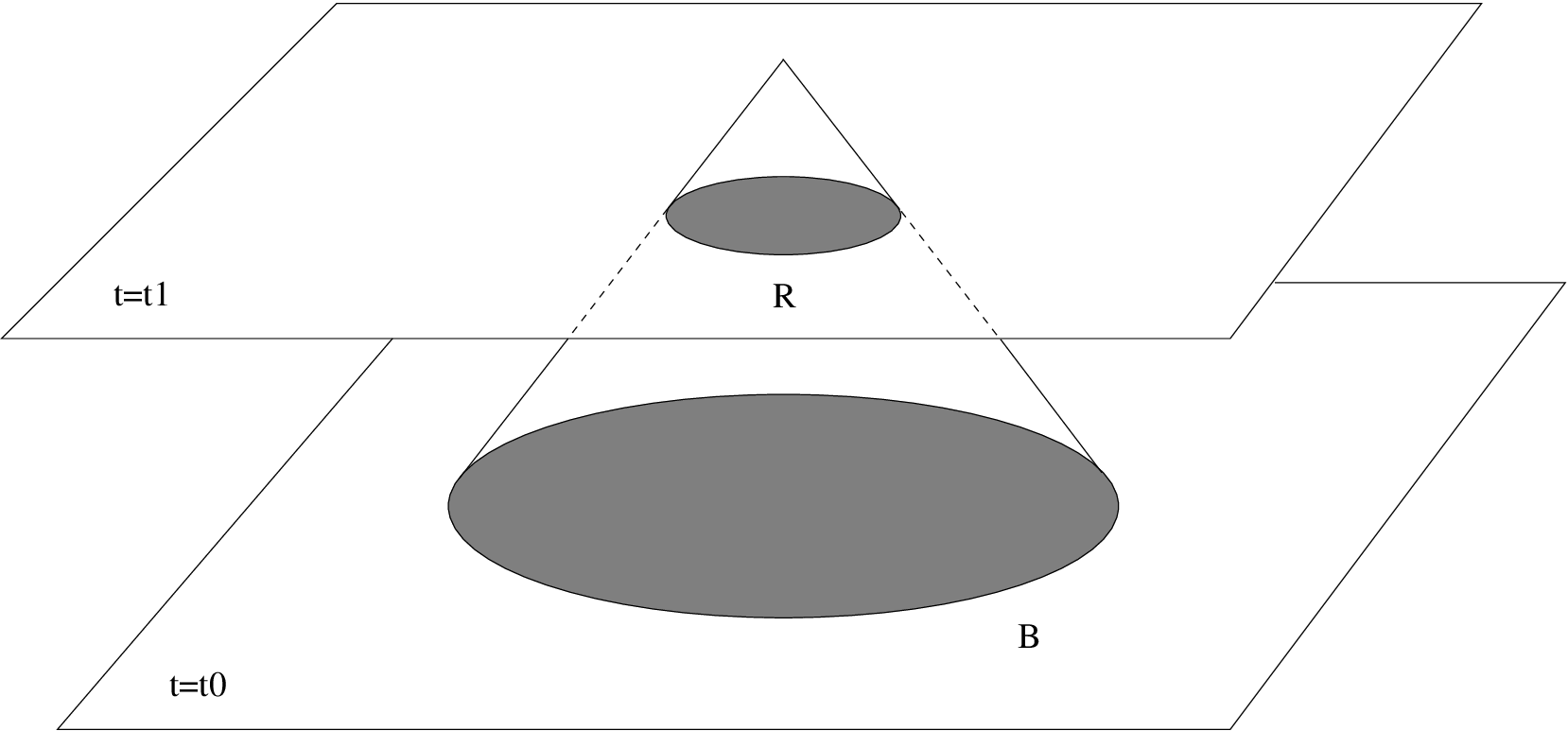}
\end{center}
\caption{Region $\mathcal{R}$ and its boundary.} \label{domain}
\end{figure}

Recall that the energy-momentum tensor for the wave equation is
\begin{equation}
T_{\mu\nu} = \partial_\mu \phi \, \partial_\nu \phi - \frac12 \partial_\alpha \phi \, \partial^\alpha \phi \, g_{\mu \nu} \, ,
\end{equation}
so that
\begin{equation}
T_{00} = \frac12 \left(\dot{\phi}^2 + a^{-2} \delta^{ij} \partial_i \phi \partial_j \phi\right) \, .
\end{equation}
Choosing the multiplier vector field
\begin{equation}
X = a^{2-n} \frac{\partial}{\partial t} \, ,
\end{equation}
we form the current
\begin{equation}
J_{\mu} = T_{\mu\nu} X^{\nu}
\end{equation}
and obtain the energy
\begin{align}
E(t) = \int_{\{t\} \times \bbR^n} J_{\mu} N^{\mu} = \int_{\bbR^n} a^2 T_{00} d^n x = \int_{\bbR^n} \frac12 \left( a^2\dot{\phi}^2 + \delta^{ij} \partial_i \phi \partial_j \phi \right) d^n x
\end{align}
(where $N = \frac{\partial}{\partial t}$ is the future unit normal). The deformation tensor associated with the multiplier $X$ is
\begin{equation}
\Pi = \frac12 \mathcal{L}_X g = - dt \mathcal{L}_X dt + \dot{a} a^{3-n} \delta_{ij} dx^i dx^j \, .
\end{equation}
Noting that
\begin{equation}
\mathcal{L}_X dt = d (\iota(X) dt) = d \left(a^{2-n}\right) = (2-n) \dot{a} a^{1-n} dt \, ,
\end{equation}
we obtain
\begin{equation}
\Pi = (n-2) \dot{a} a^{1-n} dt^2 + \dot{a} a^{3-n} \delta_{ij} dx^i dx^j \, .
\end{equation}
Therefore the bulk term is
\begin{align}
\nabla_\mu J^\mu = T^{\mu\nu} \Pi_{\mu\nu} = & \,\,(n-2) \dot{a} a^{1-n} \dot{\phi}^2 + \frac{n-2}2 \dot{a} a^{1-n} \partial_\alpha \phi \, \partial^\alpha \phi \nonumber \\
& + \dot{a} a^{-1-n} \delta^{ij} \partial_i \phi \partial_j \phi - \frac{n}2 \dot{a} a^{1-n} \partial_\alpha \phi \, \partial^\alpha \phi \\
= & \,\, (n-1) \dot{a} a^{1-n} \dot{\phi}^2 \geq 0 \, . \nonumber
\end{align}
For each $R>0$ define the set
\begin{equation}
\mathcal{B} = \left\{ (t_0,x^1, \ldots, x^n) \in I \times \bbR : \delta_{ij} x^i x^j \leq R^2 \right\} \, .
\end{equation}
Applying the divergence theorem to the current $J$ on the region
\begin{equation}
\mathcal{R} = D^+(\mathcal{B}) \cap \{ t \leq t_1 \}
\end{equation}
(see Figure~\eqref{domain}), noticing that the flux across the future null boundaries is non-positive, and letting $R \to +\infty$, we obtain

\begin{equation}
E(t_1)\leq E(t_0) < +\infty
\end{equation}
(by \eqref{finiteSobolev1}, since $k > 0$). We deduce from this inequality that, for all $t \geq t_0$,
\begin{equation}\label{energyboundF}
\int_{\bbR^n} \dot{\phi}^2 \,d^n x \lesssim \frac1{a^2}
\end{equation}
and
\begin{equation}\label{energyboundspatial}
\int_{\bbR^n} \delta^{ij} \partial_i \phi \partial_j \phi \,d^n x \lesssim 1 \, .
\end{equation}
%
%
\subsection{Partial energy}\label{subsection1.3}
Let us define the partial energy
\begin{equation}
F(t) = \int_{\bbR^n} \dot{\phi}^2 \, d^n x \, .
\end{equation}
We already know from \eqref{energyboundF} that $F \lesssim a^{-2}$,  but we want a better estimate. Differentiating this partial energy  and using the wave equation~\eqref{waveqn} gives
\begin{align}
\dot{F} & = \int_{\bbR^n} 2\dot{\phi}\ddot{\phi} \,d^n x = 2 \int_{\bbR^n} \left( -\frac{n\dot{a}}{a} \dot{\phi}^2 + \frac1{a^2} \dot{\phi} \delta^{ij} \partial_i \partial_j \phi \right) d^n x \nonumber \\
& = -\frac{2n\dot{a}}{a} F + \frac{2}{a^2} \int_{\bbR^n} \dot{\phi} \delta^{ij} \partial_i \partial_j \phi \,d^n x \, .
\end{align}
Therefore,
\begin{equation}
a^{2n}\dot{F} + 2na^{2n-1}\dot{a} F = 2a^{2n-2} \int_{\bbR^n} \dot{\phi} \delta^{ij} \partial_i \partial_j \phi \,d^n x \, ,
\end{equation}
and so, by integrating from $t_0$ to $t_1$,
\begin{equation}
F(t_1) = \frac{{a_0}^{2n}}{a(t_1)^{2n}}F_0 + \frac2{a(t_1)^{2n}} \int_{t_0}^{t_1} a^{2n-2} \int_{\bbR^n} \dot{\phi} \delta^{ij} \partial_i \partial_j \phi \,d^n x dt \, ,
\end{equation}
where $a_0=a(t_0)$ and $F_0=F(t_0)$. From the Cauchy-Schwarz inequality we have
\begin{equation}
F(t_1) \lesssim \frac1{a(t_1)^{2n}} + \frac1{a(t_1)^{2n}} \int_{t_0}^{t_1} a^{2n-2} \left(\int_{\bbR^n} \dot{\phi}^2  \,d^n x \right)^\frac12 \left(\int_{\bbR^n} (\delta^{ij} \partial_i \partial_j \phi)^2 \,d^n x \right)^\frac12 dt \, .
\end{equation}
Since each partial derivative $\partial_i\phi$ is also a solution of the wave equation, and $k \geq 2$ in \eqref{finiteSobolev1}, we have from \eqref{energyboundspatial}, applied to the partial derivatives $\partial_i\phi$, that the last integral above is bounded, whence
\begin{equation}\label{estimateF}
F(t_1) \lesssim \frac1{a(t_1)^{2n}} + \frac1{a(t_1)^{2n}} \int_{t_0}^{t_1} a^{2n-2} F^\frac12 dt \, .
\end{equation}
%
%
\subsection{Iteration}\label{subsection1.4}
Let $\varepsilon>0$ be such that
\begin{equation}
\int_{t_0}^{+\infty} a^{-\varepsilon} dt < +\infty \, ,
\end{equation}
and define
\begin{equation}
x_k = \frac{2^{k+2}-2}{2^k} - \varepsilon \frac{2^{k+1}-2}{2^k} \, .
\end{equation}
We will prove by induction that
\begin{equation}\label{estimateF2}
F \lesssim a^{-x_k}
\end{equation}
for all $k\in\mathbb{N}_0$. If $k=0$ then this is just $F\lesssim a^{-2}$, which we already had from the energy estimate. Assuming that it is true for a given $k\in\mathbb{N}_0$, we have from \eqref{estimateF} that
\begin{align}
F(t_1) & \lesssim \frac1{a(t_1)^{2n}} + \frac1{a(t_1)^{2n}} \int_{t_0}^{t_1} a^{2n-2} a^{-\frac{x_k}2} dt \nonumber \\
& \lesssim \frac1{a(t_1)^{2n}} + \frac{a(t_1)^{2n-2-\frac{x_k}2+\varepsilon}}{a(t_1)^{2n}} \int_{t_0}^{t_1} a^{-\varepsilon} dt \\
& \lesssim a(t_1)^{-\frac{x_k}2-2+\varepsilon} \nonumber
\end{align}
(where we used $x_k < 4$, so that the exponent inside the first integral is positive). Since
\begin{align}
\frac{x_k}2+2-\varepsilon & = \frac{2^{k+2}-2}{2^{k+1}} - \varepsilon \frac{2^{k+1}-2}{2^{k+1}}+2-\varepsilon \nonumber \\
& = \frac{2^{k+3}-2}{2^{k+1}} - \varepsilon \frac{2^{k+2}-2}{2^{k+1}} = x_{k+1} \, ,
\end{align}
we have established \eqref{estimateF2}.

Note that, because
\begin{equation}
\lim_{k\to+\infty} x_k = 4 - 2\varepsilon \, ,
\end{equation}
we have in fact shown that
\begin{equation}
F \lesssim a^{-4+2\varepsilon+2\delta}
\end{equation}
for any $\delta > 0$. In other words,
\begin{equation}
\| \dot{\phi} \|_{L^2(\bbR^n)} \lesssim a^{-2+\varepsilon+\delta} \, .
\end{equation}
Since any partial derivative $\partial_{i_1} \cdots \partial_{i_k} \phi$ is also a solution of the wave equation, and since \eqref{finiteSobolev1} holds, we have
\begin{equation}
\| \dot{\phi} \|_{H^k(\bbR^n)} \lesssim a^{-2+\varepsilon+\delta}
\end{equation}
for some $k > \frac{n}2$ (recall that we need one extra derivative to obtain estimate \eqref{estimateF}). Therefore, Sobolev's embedding theorem gives
\begin{equation}
| \dot{\phi} | \lesssim a^{-2+\varepsilon+\delta} \, .
\end{equation}
%
%
\section{Decay in RNdS: Proof of Theorem~\ref{mainthm2}}\label{section2}
In this section we present the proof of Theorem~\ref{mainthm2}. For the reader's convenience we break it up into elementary steps.
%
%
\subsection{Reissner-Nordstr\"{o}m-de Sitter metric}\label{subsection2.1}
The Reissner-Nordstr\"{o}m-de Sitter metric is a solution of the Einstein-Maxwell equations with positive cosmological constant, representing a pair of antipodal charged black holes in a spherical universe undergoing accelerated expansion. It is given in $n+1$ dimensions by the metric
\begin{equation} \label{RNmetric}
g = - V^{-1} dr^2 + V dt^2 + r^2 d\Omega^2 \, ,
\end{equation}
where
\begin{equation}
V = r^2 + \frac{2M}{r^{n-2}} - \frac{e^2}{r^{n-1}} - 1 \, ,
\end{equation}
and where $d\Omega^2$ is the unit round metric on $S^{n-1}$. The constants $M$ and $e$ are proportional to the mass and charge of the black holes, and we have set the cosmological constant equal to $\frac12 n (n-1)$ by an appropriate choice of units.

In the cosmological region, corresponding to $r>r_c$, we have $V>0$, and the hypersurfaces of constant $r$ are spacelike cylinders with future-pointing unit normal
\begin{equation}
N = V^\frac12 \frac{\partial}{\partial r}
\end{equation}
and volume element
\begin{equation}
dV_n = V^\frac12 r^{n-1} dt d\Omega \, .
\end{equation}
%
%
%
\subsection{Energy}\label{subsection2.2}
Assume that $\phi$ is a solution of the wave equation. Recall once again that the energy-momentum tensor associated to $\phi$ is
\begin{equation}
T_{\mu\nu} = \partial_\mu \phi \, \partial_\nu \phi - \frac12 \partial_\alpha \phi \, \partial^\alpha \phi \, g_{\mu \nu} \, .
\end{equation}
Therefore we have
\begin{align}
T(N,N) & = (N \cdot \phi)^2 + \frac12 \left[ - V \phi'^2 + V^{-1} \dot{\phi}^2 + \frac1{r^2} |\mathring{\slashed{\nabla}}\phi|^2 \right] \nonumber \\
& = \frac12 \left[ V \phi'^2 + V^{-1} \dot{\phi}^2 + \frac1{r^2} |\mathring{\slashed{\nabla}}\phi|^2 \right] \, ,
\end{align}
where $\phi' = \frac{\partial \phi}{\partial r}$, $\dot{\phi} = \frac{\partial \phi}{\partial t}$, $\mathring{\slashed{\nabla}}\phi$ is the gradient of $\phi$ seen as a function on $S^{n-1}$ and $|\mathring{\slashed{\nabla}}\phi|^2$ is its squared norm (both taken with respect to the unit round metric).

Choosing the multiplier
\begin{equation}
X = \frac{V^\frac12}{r^{n-1}} N = \frac{V}{r^{n-1}} \frac{\partial}{\partial r} \, ,
\end{equation}
we form the current
\begin{equation}
J_{\mu} = T_{\mu\nu} X^{\nu}
\end{equation}
and obtain the energy
\begin{equation}
E(r) = \int_{\bbR \times S^{n-1}} T(X,N) dV_n = \int_{\bbR \times S^{n-1}} \frac12 \left[ V^2 \phi'^2 + \dot{\phi}^2 + \frac{V}{r^2} |\mathring{\slashed{\nabla}}\phi|^2 \right] dt d\Omega \, .
\end{equation}
This energy is related to the one used by Schlue in~\cite{Schlue}, but differs (essentially) by a factor of $r$, so that no rescaling is needed at $\mathscr{I}^+$.
We will show in Section~\ref{subsection2.6} that hypotheses \eqref{hypo} imply that
\begin{equation}
\| \phi \|_{H^k(\{r=r_0\})} < + \infty
\end{equation}
for $r_0>r_c$ and $k>\frac{n}2+2>0$. In particular, $E(r_0)<+\infty$ for any $r_0>r_c$.

The deformation tensor associated to the multiplier $X$ is
\begin{equation}
\Pi = \frac12 \mathcal{L}_X g = - V^{-1} dr \mathcal{L}_X dr + \frac{V'}{2Vr^{n-1}} dr^2 + \frac{VV'}{2r^{n-1}} dt^2 + \frac{V}{r^{n-2}} d\Omega^2 \, .
\end{equation}
Noting that
\begin{equation}
\mathcal{L}_X dr = d (\iota(X) dr) = d \left(\frac{V}{r^{n-1}}\right) = \left(\frac{V'}{r^{n-1}} - \frac{(n-1)V}{r^n}\right) dr \, ,
\end{equation}
we obtain
\begin{equation}
\Pi = \frac{V'}{2r^{n-1}} \left(- V^{-1} dr^2 + V dt^2\right) + \frac{n-1}{r^n} dr^2 + \frac{V}{r^{n-2}} d\Omega^2 \, .
\end{equation}
If we write
\begin{equation}
\Pi = \Pi^{(1)} + \Pi^{(2)} + \Pi^{(3)}
\end{equation}
for each of the three terms above, we have
\begin{align}
T^{\mu\nu} \Pi^{(1)}_{\mu\nu} & = \frac{V'}{2r^{n-1}} \left[-V \phi'^2 + V^{-1} \dot{\phi}^2 - \frac22 \left( -V \phi'^2 + V^{-1} \dot{\phi}^2 + \frac1{r^2} |\mathring{\slashed{\nabla}}\phi|^2 \right) \right] \nonumber \\
& = - \frac{V'}{2r^{n+1}} |\mathring{\slashed{\nabla}}\phi|^2
\end{align}
for the first term,
\begin{align}
T^{\mu\nu} \Pi^{(2)}_{\mu\nu} & = \frac{n-1}{r^n} \left[ V^2 \phi'^2 + \frac12 V \left( -V \phi'^2 + V^{-1} \dot{\phi}^2 + \frac1{r^2} |\mathring{\slashed{\nabla}}\phi|^2 \right) \right] \nonumber \\
& = \frac{n-1}{2r^n} \left( V^2 \phi'^2 + \dot{\phi}^2 + \frac{V}{r^2} |\mathring{\slashed{\nabla}}\phi|^2 \right)
\end{align}
for the second term, and
\begin{align}
T^{\mu\nu} \Pi^{(3)}_{\mu\nu} & = \frac{V}{r^n} \left[ \frac1{r^2} |\mathring{\slashed{\nabla}}\phi|^2 - \frac{n-1}2 \left( -V \phi'^2 + V^{-1} \dot{\phi}^2 + \frac1{r^2} |\mathring{\slashed{\nabla}}\phi|^2 \right) \right] \nonumber \\
& = \frac{V}{r^{n+2}} |\mathring{\slashed{\nabla}}\phi|^2 + \frac{n-1}{2r^n} \left( V^2 \phi'^2 - \dot{\phi}^2 - \frac{V}{r^2} |\mathring{\slashed{\nabla}}\phi|^2 \right) \, .
\end{align}
for the third term. The full bulk term is therefore
\begin{equation}
\nabla_\mu J^\mu = T^{\mu\nu} \Pi_{\mu\nu} = \frac{(n-1)V^2}{r^n} \phi'^2 + \left( \frac{V}{r^{n+2}} - \frac{V'}{2r^{n+1}} \right) |\mathring{\slashed{\nabla}}\phi|^2 \, .
\end{equation}
Now,
\begin{align}
\frac{V}{r^{n+2}} - \frac{V'}{2r^{n+1}} & = \frac{1}{2r^{n-1}} \left( \frac{2V}{r^3} - \frac{V'}{r^2} \right) = - \frac{1}{2r^{n-1}} \left( \frac{V}{r^2} \right)' \nonumber \\
& = - \frac{1}{r^{n+2}} \left( 1 - \frac{nM}{r^{n-2}} + \frac{(n+1)e^2}{2r^{n-1}} \right) \geq - \frac{C}{r^{n+2}}
\end{align}
on the cosmological region $r>r_c$, and so
\begin{equation}
T^{\mu\nu} \Pi_{\mu\nu} \geq - \frac{C}{r^{n+2}} |\mathring{\slashed{\nabla}}\phi|^2 \, .
\end{equation}

\begin{figure}[h!]
\begin{center}
\psfrag{N}{$N$}
\psfrag{r=r0}{$r=r_0$}
\psfrag{r=r1}{$r=r_1$}
\psfrag{i+}{$i^+$}
\psfrag{I+}{$\mathscr{I}^+$}
\psfrag{CH+1}{$\mathcal{CH}_1^+$}
\psfrag{CH+2}{$\mathcal{CH}_2^+$}
\psfrag{C}{$\mathcal{C}$}
\psfrag{S}{$\mathcal{S}$}
\epsfxsize=0.8\textwidth
\leavevmode
\epsfbox{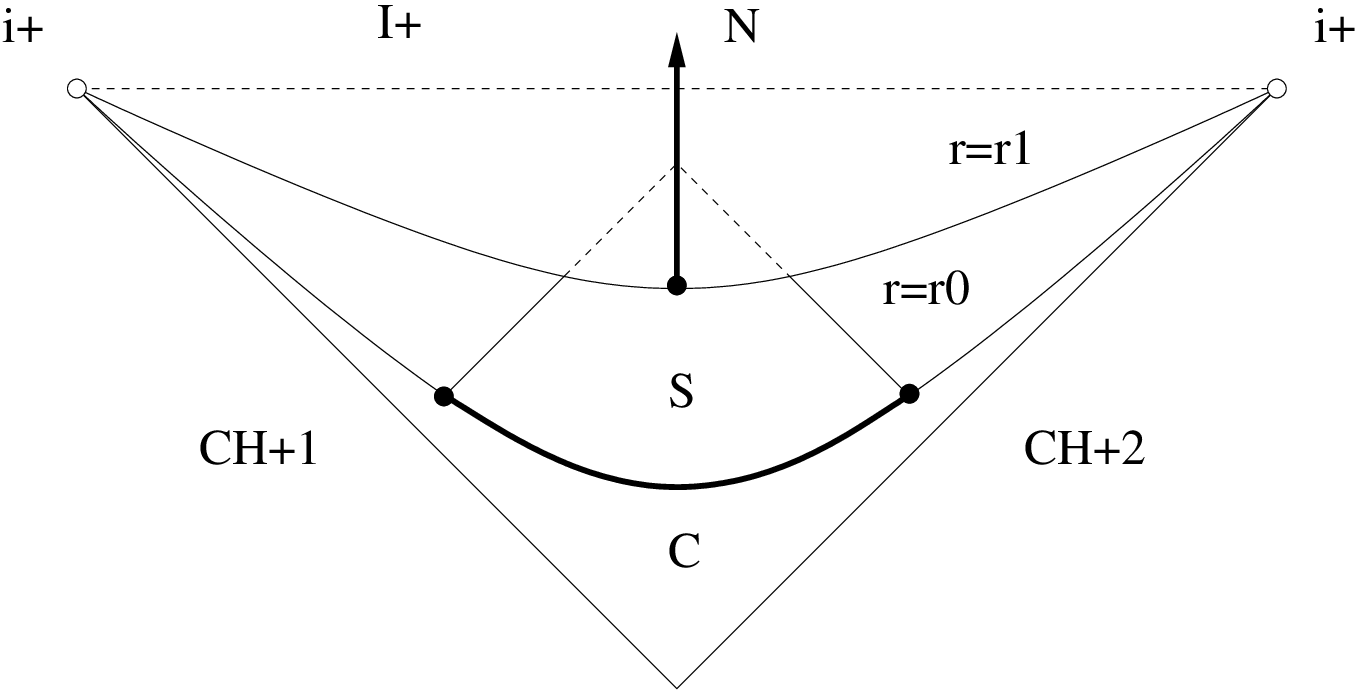}
\end{center}
\caption{Region $\mathcal{S}$ and its boundary.} \label{domain2}
\end{figure}

For each $T>0$ define the set
\begin{equation}
\mathcal{C} = \{ r = r_0 \} \cap \{ -T \leq t \leq T\} \, .
\end{equation}
Applying the divergence theorem to the current $J$ on the region
\begin{equation}
\mathcal{S} = D^+(\mathcal{C}) \cap \{ r \leq r_1 \}
\end{equation}
(see Figure~\eqref{domain2}), noticing that the flux across the future null boundaries is non-positive, and letting $T \to +\infty$, we obtain

\begin{equation}
E(r_0) - E(r_1) \geq - \int_{r_0}^{r_1} \int_{\bbR \times S^{n-1}} \frac{C}{r^{n+2}} |\mathring{\slashed{\nabla}}\phi|^2 r^{n-1} dtd\Omega dr \, .
\end{equation}
Notice that, for $r \geq r_0$,
\begin{equation}
\int_{\bbR \times S^{n-1}} |\mathring{\slashed{\nabla}}\phi|^2 dt d\Omega \leq \frac{2r^2}{V} E(r) \leq C(r_0) E(r) \, ,
\end{equation}
since $\frac{2r^2}{V}$ is positive for $r>r_c$ and
\begin{equation}
\lim_{r \to \infty} \frac{2r^2}{V} = 2 \, .
\end{equation}
Substituting in the previous inequality yields
\begin{equation}
E(r_1) \leq E(r_0) + \int_{r_0}^{r_1} \frac{C(r_0)}{r^3} E(r) dr \, .
\end{equation}
From Gr\"{o}nwall's inequality we finally obtain
\begin{equation}
E(r_1) \leq E(r_0) \exp\left(\int_{r_0}^{r_1} \frac{C(r_0)}{r^3} dr\right) \leq C(r_0) E(r_0) \, .
\end{equation}
In particular, we have, for all $r \geq r_0$:
\begin{align}
& \int_{\bbR \times S^{n-1}} V^2 \phi'^2 dt d\Omega \lesssim 1 \, ; \label{energyboundF2} \\
& \int_{\bbR \times S^{n-1}} \dot{\phi}^2 dt d\Omega \lesssim 1 \, ; \label{energyboundt} \\
& \int_{\bbR \times S^{n-1}} \frac{V}{r^2} |\mathring{\slashed{\nabla}}\phi|^2 dt d\Omega \lesssim 1 \, . \label{energyboundlap}
\end{align}
%
%
\subsection{Wave equation}\label{subsection2.3}
The wave equation in the RNdS background,
\begin{equation}
\Box_g \phi = 0 \Leftrightarrow \partial_\mu \left(\sqrt{-g} \, \partial^\mu \phi \right) = 0 \Leftrightarrow \partial_\mu \left(r^{n-1} \left(\mathring{\slashed{g}}\right)^\frac12 \partial^\mu \phi \right) = 0
\end{equation}
(where $\mathring{\slashed{g}}$ is the determinant of the unit round sphere metric), can be written as
\begin{align} \label{waveqnRN}
& - \left( r^{n-1} V \phi' \right)' + r^{n-1} V^{-1} \ddot{\phi} + r^{n-3} \mathring{\slashed{\Delta}} \phi = 0 \Leftrightarrow \nonumber \\
& - \left( V \phi' \right)' - \frac{n-1}{r} V \phi' + V^{-1} \ddot{\phi} + \frac1{r^2} \mathring{\slashed{\Delta}} \phi = 0 \, ,
\end{align}
where $\mathring{\slashed{\Delta}}\phi$ is the Laplacian of $\phi$ seen as a function on $S^{n-1}$ (taken with respect to the unit round metric).
%
%
\subsection{Partial energy}\label{subsection2.4}
Let us define the partial energy
\begin{equation}
F(r) = \int_{\bbR \times S^{n-1}} V^2 \phi'^2 dt d\Omega \, .
\end{equation}
We already know from \eqref{energyboundF2} that $F \lesssim 1$,  but we want a better estimate. Differentiating this partial energy and using the wave equation gives
\begin{align}
F' & = 2 \int_{\bbR \times S^{n-1}} V \phi' (V \phi')' dt d\Omega \nonumber  \\
& = 2 \int_{\bbR \times S^{n-1}} V \phi' \left( - \frac{n-1}{r} V \phi' + V^{-1} \ddot{\phi} + \frac1{r^2} \mathring{\slashed{\Delta}} \phi \right) dt d\Omega \label{RNFeqn} \\
& = - \frac{2n-2}{r} F + 2 \int_{\bbR \times S^{n-1}} V \phi' \left( V^{-1} \ddot{\phi} + \frac1{r^2} \mathring{\slashed{\Delta}} \, \phi \right) dt d\Omega \, . \nonumber
\end{align}
Noting that
\begin{equation}
F' + \frac{2n-2}{r} F = \frac1{r^{2n-2}}\left(r^{2n-2} F\right)' \, ,
\end{equation}
we can integrate \eqref{RNFeqn} to obtain
\begin{align}
F(r_1) & = \frac{{r_0}^{2n-2}}{{r_1}^{2n-2}} F(r_0) + \frac{2}{{r_1}^{2n-2}} \int_{r_0}^{r_1} \int_{\bbR \times S^{n-1}} r^{2n-2} V \phi' \left( V^{-1} \ddot{\phi} + \frac1{r^2} \mathring{\slashed{\Delta}} \phi \right) dt d\Omega dr \nonumber \\
& \leq \frac{{r_0}^{2n-2}}{{r_1}^{2n-2}} F(r_0) + \frac{2\sqrt{2}}{{r_1}^{2n-2}} \int_{r_0}^{r_1} r^{2n-4} F^\frac12 \left(\int_{\bbR \times S^{n-1}} \left( \frac{r^4}{V^2} \ddot{\phi}^2 + \left(\mathring{\slashed{\Delta}} \phi \right)^2 \right) dt d\Omega \right)^\frac12 dr \, , \label{FleqF}
\end{align}
where we used the Cauchy-Schwarz inequality in the last step.

Recall that $S^{n-1}$ admits $\frac{n(n-1)}2$ independent Killing vector fields, given by
\begin{equation}
L_{ij} = x^i \frac{\partial}{\partial x^j} - x^j \frac{\partial}{\partial x^i}
\end{equation}
for $i<j$ (under the usual embedding $S^{n-1} \subset \bbR^n$), and moreover that
\begin{equation}
\mathring{\slashed{\Delta}} \phi = \sum_{i<j} L_{ij} \cdot (L_{ij} \cdot \phi) \, .
\end{equation}
Since $\frac{\partial}{\partial t}$ and $L_{ij}$ are Killing vector fields, $\dot{\phi}$ and $L_{ij} \cdot \phi$ are also solutions of the wave equation, and, because $k \geq 2$ in \eqref{hypo}, they satisfy the bounds \eqref{energyboundt} and \eqref{energyboundlap}. Using
\begin{equation}
(L_{ij} \cdot (L_{ij} \cdot \phi))^2 \leq (\mathring{\slashed{\nabla}} (L_{ij} \cdot \phi))^2 \, ,
\end{equation}
we see that the last integral in \eqref{FleqF} is bounded, whence
\begin{equation}\label{estimateFRN}
F(r_1) \lesssim \frac1{{r_1}^{2n-2}} + \frac1{{r_1}^{2n-2}} \int_{r_0}^{r_1} r^{2n-4} F^\frac12 dr \, .
\end{equation}
%
%
\subsection{Iteration}\label{subsection2.5}
Define
\begin{equation}
x_k = \frac{2^{k+1}-2}{2^k} \, .
\end{equation}
We will prove by induction that
\begin{equation}\label{estimateF2RN}
F(r_1) \lesssim {r_1}^{-x_k}
\end{equation}
for all $k\in\mathbb{N}_0$. If $k=0$ then this is just $F\lesssim 1$, which we already had from the energy estimate. Assuming that it is true for a given $k\in\mathbb{N}_0$, we have from \eqref{estimateFRN} that
\begin{align}
F(r_1) & \lesssim \frac1{{r_1}^{2n-2}} + \frac1{{r_1}^{2n-2}} \int_{r_0}^{r_1} r^{2n-4} r^{-\frac{x_k}2} dr \lesssim \frac1{{r_1}^{2n-2}} + \frac{{r_1}^{2n-3-\frac{x_k}2}}{{r_1}^{2n-2}} \nonumber \\
& \lesssim {r_1}^{-\frac{x_k}2-1}
\end{align}
(where we used $x_k < 2$, so that the exponent $2n-3-\frac{x_k}2$ is positive). Since
\begin{equation}
\frac{x_k}2+1 = \frac{2^{k+1}-2}{2^{k+1}} + 1 = \frac{2^{k+2}-2}{2^{k+1}} = x_{k+1} \, ,
\end{equation}
we have established \eqref{estimateF2RN}.

Note that, since
\begin{equation}
\lim_{k\to+\infty} x_k = 2 \, ,
\end{equation}
we have shown that
\begin{equation}
F(r_1) \lesssim {r_1}^{-2+2\delta}
\end{equation}
for any $\delta > 0$. Noticing that
\begin{equation}
V(r_1) \sim {r_1}^2 \, ,
\end{equation}
we see that, in fact,
\begin{equation}
\| \phi' \|_{L^2(\bbR \times S^{n-1})} \lesssim {r_1}^{-3+\delta} \, .
\end{equation}
Since we obtain solutions of the wave equation by acting on $\phi$ with any finite sequence of Killing vector fields $\frac{\partial}{\partial t}$ and $L_{ij}$, and since \eqref{hypo} holds, we have
\begin{equation}
\| \phi' \|_{H^k(\bbR \times S^{n-1})} \lesssim {r_1}^{-3+\delta}
\end{equation}
for some $k > \frac{n}2$ (recall that we need one extra derivative to obtain estimate \eqref{estimateFRN}). Therefore Sobolev's embedding theorem gives\footnote{Sobolev's embedding theorem holds for any complete Riemannian manifold with positive injectivity radius and bounded sectional curvature, see for instance \cite{Aubin}.}
\begin{equation}
| \phi' | \lesssim {r_1}^{-3+\delta} \, .
\end{equation}
%
%
%
\subsection{Weak redshift estimates}\label{subsection2.6}
\begin{figure}[h!]
\begin{center}
\psfrag{r=const.}{$r=\,\,$const.}
\psfrag{u=const.}{$u=\,\,$const.}
\psfrag{v=const.}{$v=\,\,$const.}
\psfrag{u=-infty}{$u=-\infty$}
\psfrag{v=-infty}{$v=-\infty$}
\psfrag{i+}{$i^+$}
\psfrag{I+}{$\mathscr{I}^+$}
\psfrag{CH+1}{$\mathcal{CH}_1^+$}
\psfrag{CH+2}{$\mathcal{CH}_2^+$}
\epsfxsize=0.8\textwidth
\leavevmode
\epsfbox{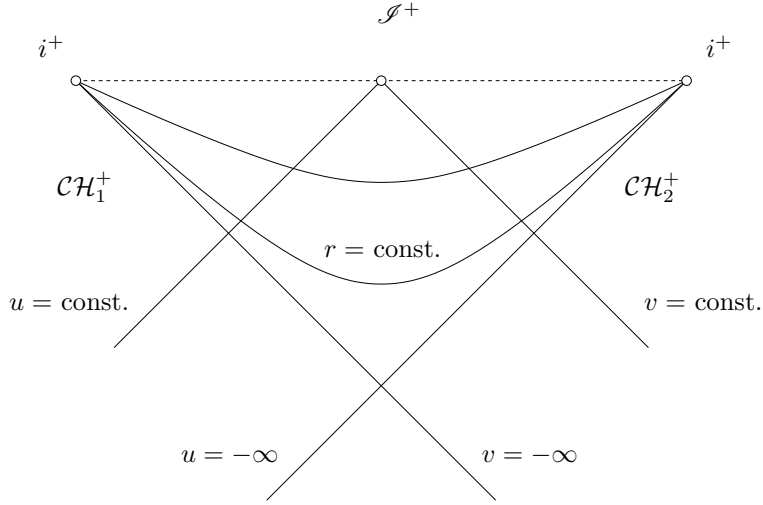}
\end{center}
\caption{Coordinates $u$ and $v$ (note that $t$ increases from right to left).} \label{domain3}
\end{figure}

We now obtain the condition that must be satisfied at the cosmological horizon so that the energy $E(r)$, corresponding to the multiplier $X$, is finite at $r=r_0$. We start by writing the Reissner-Nordstr\"{o}m-de Sitter metric \eqref{RNmetric} as
\begin{align}
g & = V \left(- V^{-2} dr^2 + dt^2\right) + r^2 d\Omega^2 \nonumber \\
& = - V \left(V^{-1} dr + dt\right) \left(V^{-1} dr - dt\right) + r^2 d\Omega^2 \nonumber \\
& = - V du \left(-du + 2V^{-1} dr\right) + r^2 d\Omega^2 \nonumber \\
& = V du^2 - 2 du dr + r^2 d\Omega^2 \, ,
\end{align}
where the coordinate $u$ is defined as
\begin{equation}
u = t + \int \frac{dr}{V} \, .
\end{equation}
The first diagonal block for the matrix of the metric in the coordinates $(u,r)$ satisfies
\begin{equation}
\det\left(
\begin{matrix}
V & -1 \\ -1 & 0
\end{matrix}
\right)
= -1 \, ,
\end{equation}
and so this coordinate system extends across the cosmological horizon $r=r_c$, where $V=0$. Note that the hypersurfaces of constant $u$ are null and transverse to the cosmological horizon, and so only one of the branches of the cosmological horizon (connecting the bifurcation sphere to future null infinity $\mathscr{I}^+$) is covered by the coordinates $(u,r)$; to cover the other branch, corresponding to $u=-\infty$, one has to introduce new coordinates $(v,r)$, defined by
\begin{equation}
v = - t + \int \frac{dr}{V} \, ,
\end{equation}
and repeat the same construction (see Figure~\ref{domain3}).

At any rate, the Killing vector field
\begin{equation}
K = \frac{\partial}{\partial u} = \frac{\partial}{\partial t}
\end{equation}
is well-defined across the (first branch of the) cosmological horizon, and is null on the cosmological horizon (although the coordinate $t$ is not defined there). Moreover, the vector field
\begin{equation}
Y = \left(\frac{\partial}{\partial r}\right)_u
\end{equation}
is null and transverse to the cosmological horizon. From the identities
\begin{equation}
du(Y) = 0  \qquad \text{ and } \qquad Y \cdot r = 1
\end{equation}
one easily obtains
\begin{equation}
Y = \frac{\partial}{\partial r} - \frac1{V} \frac{\partial}{\partial t}
\end{equation}
on the cosmological region. Finally, to find the expression for the multiplier vector field $X$ in the coordinates $(u,r)$, we start by computing
\begin{equation}
N = - \frac{\grad r}{|\grad r|} \, .
\end{equation}
Given that
\begin{equation}
\left(
\begin{matrix}
V & -1 \\ -1 & 0
\end{matrix}
\right)^{-1}
=
\left(
\begin{matrix}
0 & -1 \\ -1 & -V
\end{matrix}
\right) \, ,
\end{equation}
we have
\begin{equation}
\left\langle \grad r, \grad r \right\rangle = \left\langle dr, dr \right\rangle = -V,
\end{equation}
and so $N$ is the vector associated to the covector $-V^{-\frac12} dr$, that is,
\begin{equation}
N = V^{-\frac12} \left(\frac{\partial}{\partial u} + V \frac{\partial}{\partial r}\right) \, .
\end{equation}
Consequently,
\begin{equation}
X = \frac{V^\frac12}{r^{n-1}} N = \frac1{r^{n-1}} \left(\frac{\partial}{\partial u} + V \frac{\partial}{\partial r}\right)
\end{equation}
is well-defined across the cosmological horizon.

Note that the energy
\begin{equation}
E(r) = \int_{\bbR \times S^{n-1}} T(X,N) dV_n = \int_{\bbR \times S^{n-1}} \frac12 \left[ V^2 \phi'^2 + \dot{\phi}^2 + \frac{V}{r^2} |\mathring{\slashed{\nabla}}\phi|^2 \right] dt d\Omega
\end{equation}
approaches
\begin{equation}
E(r_c) = \frac12 \int_{\bbR \times S^{n-1}} (K \cdot \phi)^2 du d\Omega + \frac12 \int_{\bbR \times S^{n-1}} (K \cdot \phi)^2 dv d\Omega
\end{equation}
as $r \to r_c$, where each of the two integrals above refers to a different branch of the cosmological horizon. It is therefore clear that $E(r)$ loses control of transverse and angular derivatives as $r \to r_c$. To cincumvent this problem, we define a new energy by adding the vector field $Y$ to the original multiplier $X$:
\begin{equation}
\mathcal{E}(r) = E(r) + \int_{\bbR \times S^{n-1}} T(Y,N) dV_n \, .
\end{equation}
Now,
\begin{align}
T(Y,N) & = T\left(\frac{\partial}{\partial r}, N\right) - \frac1{V} T\left(\frac{\partial}{\partial t}, N\right) \nonumber \\
& = V^{-\frac12} \left[T(N,N) - T\left(\frac{\partial}{\partial t}, \frac{\partial}{\partial r}\right)\right] \, ,
\end{align}
and so
\begin{align}
\mathcal{E}(r) & = E(r) + \int_{\bbR \times S^{n-1}} \frac12 \left[ V \phi'^2 + V^{-1} \dot{\phi}^2 + \frac1{r^2} |\mathring{\slashed{\nabla}}\phi|^2 - 2\dot{\phi} \phi' \right] r^{n-1}dt d\Omega \nonumber \\
& = E(r) + \int_{\bbR \times S^{n-1}} \frac12 \left[ V \left( \phi' - \frac1{V} \dot{\phi} \right)^2 + \frac1{r^2} |\mathring{\slashed{\nabla}}\phi|^2 \right] r^{n-1}dt d\Omega \nonumber \\
& = E(r) + \int_{\bbR \times S^{n-1}} \frac12 \left[ V \left( Y \cdot \phi \right)^2 + \frac1{r^2} |\mathring{\slashed{\nabla}}\phi|^2 \right] r^{n-1}dt d\Omega \, .
\end{align}
Note that
\begin{equation}
\mathcal{E}(r_c) = E(r_c) + \frac{{r_c}^{n-3}}2 \int_{\bbR \times S^{n-1}} |\mathring{\slashed{\nabla}}\phi|^2  du d\Omega + \frac{{r_c}^{n-3}}2 \int_{\bbR \times S^{n-1}} |\mathring{\slashed{\nabla}}\phi|^2  dv d\Omega \, ,
\end{equation}
and so the new energy retains some control of the angular derivatives as $r \to r_c$. Note that this energy is a weaker version of the Dafermos-Rodnianski redshift energy, which also controls transverse derivatives. Nevertheless, our (simpler) construction suffices to show that $E(r_0)$ is finite from hypotheses~\eqref{hypo}.

To compute the deformation tensor associated to the multiplier $Y$ we note that
\begin{equation}
\mathcal{L}_{\frac{\partial}{\partial r}} g = V^{-2}V' dr^2 + V' dt^2 + 2r d\Omega^2
\end{equation}
and
\begin{equation}
\mathcal{L}_{-\frac1{V}\frac{\partial}{\partial t}} g = 2Vdt \mathcal{L}_{-\frac1{V}\frac{\partial}{\partial t}} dt = 2Vdt d\left(-\frac1{V}\right) = 2V^{-1}V' dt dr \, .
\end{equation}
Therefore, the deformation tensor is
\begin{align}
\Xi & = \frac12 \mathcal{L}_{Y} g = \frac12 V^{-2}V' dr^2 + \frac12 V' dt^2 + r d\Omega^2 + V^{-1}V' dt dr \nonumber  \\
& = \frac12 V' \left( dt + V^{-1} dr \right)^2  + r d\Omega^2 = \frac12 V' du^2  + r d\Omega^2 \, .
\end{align}
Noticing that
\begin{equation}
du = - g(Y,\cdot) \, ,
\end{equation}
we have
\begin{equation}
T^{\mu\nu} \Xi_{\mu\nu} = \frac12 V' (Y \cdot \phi)^2 + \frac1{r^3} |\mathring{\slashed{\nabla}}\phi|^2 - \frac{n-1}{2r} \left\langle d\phi, d\phi \right\rangle \, .
\end{equation}
Since
\begin{equation}
\left\langle d\phi, d\phi \right\rangle = -2(K \cdot \phi)(Y \cdot \phi) - V(Y \cdot \phi)^2 + \frac1{r^2} |\mathring{\slashed{\nabla}}\phi|^2 \, ,
\end{equation}
we finally obtain
\begin{equation}
T^{\mu\nu} \Xi_{\mu\nu} = \left(\frac{V'}2 + \frac{(n-1)V}{2r} \right)(Y \cdot \phi)^2 + \frac{n-1}{r}(K \cdot \phi)(Y \cdot \phi) - \frac{n-3}{2r^3} |\mathring{\slashed{\nabla}}\phi|^2 \, .
\end{equation}
Since
\begin{equation}
\frac{V'}2 (Y \cdot \phi)^2 + \frac{n-1}{r}(K \cdot \phi)(Y \cdot \phi) = \frac{V'}2 \left[ (Y \cdot \phi) + \frac{n-1}{rV'}(K \cdot \phi)\right]^2 - \frac{(n-1)^2}{2r^2V'}(K \cdot \phi)^2 \, ,
\end{equation}
and using the fact that $V'(r)>0$ for $r\geq r_c$ (global redshift), we have
\begin{equation}
T^{\mu\nu} \Xi_{\mu\nu} \geq - \frac{(n-1)^2}{2r^2V'}(K \cdot \phi)^2 - \frac{n-3}{2r^3} |\mathring{\slashed{\nabla}}\phi|^2 \, ,
\end{equation}
and so
\begin{equation}
T^{\mu\nu} \Pi_{\mu\nu} + T^{\mu\nu} \Xi_{\mu\nu} \geq - C (K \cdot \phi)^2 - C |\mathring{\slashed{\nabla}}\phi|^2
\end{equation}
for $r_c < r < r_0$.

Given $r_c < r _1 < r_0$ and $T>0$, define the set
\begin{equation}
\mathcal{D} = \{ r = r_1 \} \cap \{ -T \leq t \leq T\} \, .
\end{equation}
Applying the divergence theorem on the region
\begin{equation}
\mathcal{T} = D^+(\mathcal{D}) \cap \{ r \leq r_0 \},
\end{equation}
noticing that the flux across the future null boundaries is non-positive, and letting $T \to +\infty$, we obtain
\begin{equation}
\mathcal{E}(r_1)- \mathcal{E}(r_0) \geq - \int_{r_1}^{r_0} \int_{\bbR \times S^{n-1}} C \left( (K \cdot \phi)^2 + |\mathring{\slashed{\nabla}}\phi|^2\right) r^{n-1} dtd\Omega dr \, .
\end{equation}
Since
\begin{align}
\mathcal{E}(r) & = \int_{\bbR \times S^{n-1}} \frac12 \left[ V^2 \phi'^2 + \left( K \cdot \phi \right)^2 + \frac{V}{r^2} |\mathring{\slashed{\nabla}}\phi|^2 \right] dt d\Omega \nonumber \\
&  + \int_{\bbR \times S^{n-1}} \frac12 \left[ V \left( Y \cdot \phi \right)^2 + \frac1{r^2} |\mathring{\slashed{\nabla}}\phi|^2 \right] r^{n-1}dt d\Omega \, ,
\end{align}
we have, for $r_c < r_1 < r_0$,
\begin{equation}
\mathcal{E}(r_1) - \mathcal{E}(r_0) \geq - \int_{r_1}^{r_0} C\mathcal{E}(r) dr \, ,
\end{equation}
that is,
\begin{equation}
\mathcal{E}(r_0) \leq \mathcal{E}(r_1) + \int_{r_1}^{r_0} C\mathcal{E}(r) dr \, .
\end{equation}
From Gr\"{o}nwall's inequality we have
\begin{equation}
\mathcal{E}(r_0) \leq \mathcal{E}(r_1)\exp\left(\int_{r_1}^{r_0} C dr \right) \leq C(r_0) \mathcal{E}(r_1) \, .
\end{equation}
Letting $r_1 \to r_c$, we finally obtain
\begin{equation}
E(r_0) \leq \mathcal{E}(r_0) \lesssim \mathcal{E}(r_c) \lesssim \| \phi \|_{H^1(\mathcal{CH}_1^+)} + \| \phi \|_{H^1(\mathcal{CH}_2^+)} < + \infty \, .
\end{equation}
Commuting with the Killing vector fields $\frac{\partial}{\partial t}$ and $L_{ij}$, we see that hypotheses \eqref{hypo} imply that
\begin{equation}
\| \phi \|_{H^k(\{r=r_0\})} \lesssim \| \phi \|_{H^k(\mathcal{CH}_1^+)} + \| \phi \|_{H^k(\mathcal{CH}_2^+)} < + \infty
\end{equation}
for some $k>\frac{n}2+2$.
%
%
\section*{Acknowledgements}
This work was partially supported by FCT/Portugal through UID/MAT/04459/2013 and grant (GPSEinstein) PTDC/MAT-ANA/1275/2014.
Pedro Oliveira was supported by FCT/Portugal through the LisMath scholarship PD/BD/52640/2014.
%
%
\appendix
%
%
\section{Fourier modes}\label{appendixA}
It should be clear from the proof of Theorem~\ref{mainthm1} that this result also holds for expanding flat FLRW models with toroidal spatial sections; this shows, in particular, that the underlying decay mechanism must be the cosmological expansion, as opposed to dispersion. In the toroidal case, the wave equation can be studied by performing a Fourier mode analysis, which gives valuable information about how sharp our estimates are.

Taking, for simplicity, $\mathbb{T}^n=\mathbb{R}^n/(2\pi\mathbb{Z})^n$, we can expand any smooth function $\phi:\mathbb{R}\times \mathbb{T}^n\to\mathbb{R}$ as
\begin{equation}
\phi(t,x)=\sum_{k\in\mathbb{Z}^n} c_k(t) e^{i\langle k,x\rangle} \, .
\end{equation}
Substituting in \eqref{waveqn} we obtain
\begin{equation} \label{modeqn}
\ddot{c}_k + \frac{n\dot{a}}{a} \dot{c}_k + \frac{k^2}{a^2} c_k = 0 \, ,
\end{equation}
or, equivalently,
\begin{equation}
 \frac{d}{dt} \left( a^n \dot{c}_k \right) + k^2 a^{n-2} c_k = 0 \, .
\end{equation}
If we change the independent variable to
\begin{equation}
\tau = \int \frac{dt}{a(t)} \, ,
\end{equation}
so that
\begin{equation}
\frac{d}{dt} = \frac1{a}\frac{d}{d\tau} \, ,
\end{equation}
equation \eqref{modeqn} becomes
\begin{equation}
\left( a^{n-1} c_k' \right)' + k^2 a^{n-1} c_k = 0 \, ,
\end{equation}
where the prime denotes differentiation with respect to $\tau$. Setting
\begin{equation}
c_k = a^{-\frac{n-1}2} d_k \, ,
\end{equation}
so that
\begin{equation}
c_k' = a^{-\frac{n-1}2} d_k'-\frac{(n-1)}2a^{-\frac{n-1}2-1} a' d_k \, ,
\end{equation}
we obtain
\begin{equation}
\left( a^{\frac{n-1}2} d_k'-\frac{(n-1)}2a^{\frac{n-1}2-1} a' d_k \right)' + k^2 a^{\frac{n-1}2} d_k = 0 \, ,
\end{equation}
or, equivalently,
\begin{equation} \label{eqndk}
d_k'' + \left[k^2 - \frac{(n-1)}{2} \frac{a''}{a} - \frac{(n-1)(n-3)}{4} \left(\frac{a'}{a}\right)^2 \right]d_k = 0 \, .
\end{equation}

If $a(t)=t^p$, then
\begin{equation}
\tau = \int \frac{dt}{t^p} = \frac{t^{1-p}}{1-p} \Leftrightarrow t = \left[(1-p)\tau\right]^\frac1{1-p} \, ,
\end{equation}
whence
\begin{equation}
a = \left[(1-p)\tau\right]^\frac{p}{1-p} \, ,
\end{equation}
thus implying
\begin{equation}
a' = p\left[(1-p)\tau\right]^{\frac{p}{1-p}-1}
\end{equation}
and
\begin{equation}
a''= p(2p-1)\left[(1-p)\tau\right]^{\frac{p}{1-p}-2} \, .
\end{equation}
We conclude that equation \eqref{eqndk} can be written as
\begin{equation} \label{eqndk2}
d_k'' + \left(k^2 - \frac{\mu}{\tau^2} \right)d_k = 0 \, ,
\end{equation}
where
\begin{equation}
\mu= \frac{(n-1)p(2p-1)}{2(1-p)^2} + \frac{(n-1)(n-3)p^2}{4(1-p)^2} \, .
\end{equation}
The general solution of equation~\eqref{eqndk2} is
\begin{equation}
d_k(\tau) = C_1 \sqrt{\tau} J_\nu (k\tau) + C_2 \sqrt{\tau} Y_{\nu} (k\tau) \, ,
\end{equation}
where $k=|k|$,
\begin{equation}
J_\alpha (z) = \sum_{m=0}^{+\infty} \frac{(-1)^m}{m!\,\Gamma(m+\alpha+1)}\left(\frac{z}{2}\right)^{2m+\alpha}
\end{equation}
is the Bessel function of the first kind,
\begin{equation}
Y_\alpha (z) = \frac{J_\alpha (z)\cos(\alpha\pi)-J_{-\alpha} (z)}{\sin(\alpha\pi)}
\end{equation}
is the Bessel function of the second kind, and
\begin{equation}
\nu^2 = \frac14 + \mu = \frac{(np-1)^2}{4(1-p)^2} \geq 0 \, .
\end{equation}

For $p>1$ we have
\begin{equation}
d_k \sim C_1 \tau^{\frac12-\nu} + C_2 \tau^{\frac52-\nu}
\end{equation}
as $t\to+\infty\Leftrightarrow \tau\to 0^-$ (note that  $\nu > \frac{n}2$ for $p>1$), whence
\begin{equation}
d_k \sim C_1 t^{(1-p)(\frac12-\nu)} + C_2 t^{(1-p)(\frac52-\nu)} \, .
\end{equation}
This leads to
\begin{equation}
c_k \sim C_1 t^{-\frac{n-1}{2}p+(1-p)(\frac12-\nu)} + C_2 t^{-\frac{n-1}{2}p+(1-p)(\frac52-\nu)} \, ,
\end{equation}
that is,
\begin{equation}
c_k \sim C_1 + C_2 t^{-2p+2} \, ,
\end{equation}
implying in particular that
\begin{equation}
|\dot{c}_k| \lesssim t^{-2p+1} \, .
\end{equation}
Comparing with \eqref{mainestimatep}, we see that \eqref{mainestimate} is (almost) sharp in the case $a(t)=t^p$ with $p>1$.

For $p<1$ we have\footnote{More precisely, as $z \to \infty$ we have the asymptotic formulae
\begin{align*}
J_\nu(z) \sim \sqrt{\frac{2}{\pi z}} \cos\left(z-\frac{\nu\pi}{2}-\frac{\pi}{4}\right), \qquad
Y_\nu(z) \sim \sqrt{\frac{2}{\pi z}} \sin\left(z-\frac{\nu\pi}{2}-\frac{\pi}{4}\right) \, ,
\end {align*}
which hold up to the first derivative, as can be seen from the identities
\begin{align*}
\frac{dJ_\nu}{dz}(z) = \frac12 J_{\nu-1}(z) - \frac12 J_{\nu+1}(z) \, , \qquad
\frac{dY_\nu}{dz}(z) = \frac12 Y_{\nu-1}(z) - \frac12 Y_{\nu+1}(z) \, .
\end {align*}
}
\begin{equation}
d_k \sim C_1\cos\left(k\tau - \frac{\nu\pi}2 - \frac{\pi}4 \right) + C_2\cos\left(k\tau + \frac{\nu\pi}2 - \frac{\pi}4 \right)
\end{equation}
as $t\to+\infty\Leftrightarrow \tau\to +\infty$, whence
\begin{equation}
d_k \sim C_1\cos\left(\frac{kt^{1-p}}{1-p} - \frac{\nu\pi}2 - \frac{\pi}4 \right) + C_2\cos\left(\frac{kt^{1-p}}{1-p} + \frac{\nu\pi}2 - \frac{\pi}4 \right) \, .
\end{equation}
This leads to
\begin{equation}
c_k \sim C_1 t^{-\frac{n-1}{2}p} \cos\left(\frac{kt^{1-p}}{1-p} - \frac{\nu\pi}2 - \frac{\pi}4 \right) + C_2 t^{-\frac{n-1}{2}p} \cos\left(\frac{kt^{1-p}}{1-p} + \frac{\nu\pi}2 - \frac{\pi}4 \right) \, ,
\end{equation}
implying in particular that
\begin{equation}
|\dot{c}_k| \lesssim t^{-\frac{n+1}{2}p} \, .
\end{equation}
Comparing with \eqref{mainestimatep}, we see that \eqref{mainestimate} is very far from sharp in the case $a(t)=t^p$ with $p<1$. Note that in this case we do obtain the exponent coming from the na{\"\i}ve physical argument governing the decay of the Fourier modes.

If $a(t)=e^t$, then\footnote{Note that we can always set $H=1$ by choosing units such that $\Lambda = \frac12 n (n-1)$.}
\begin{equation}
\tau = \int \frac{dt}{e^t} = -e^{-t} \Leftrightarrow t = -\log(-\tau) \, ,
\end{equation}
whence
\begin{equation}
a = - \frac1{\tau} \, ,
\end{equation}
implying
\begin{equation}
a' = \frac1{\tau^2}
\end{equation}
and
\begin{equation}
a''= -\frac2{\tau^3} \, .
\end{equation}
We conclude that equation \eqref{eqndk} can be written as
\begin{equation} \label{eqndk3}
d_k'' + \left(k^2 - \frac{\mu}{\tau^2} \right)d_k = 0 \, ,
\end{equation}
where
\begin{equation}
\mu= n-1 + \frac{(n-1)(n-3)}{4}
\end{equation}
(the limit of the value in the $a(t)=t^p$ case as $p \to +\infty$). The general solution of equation~\eqref{eqndk3} is
\begin{equation}
d_k(\tau) = C_1 \sqrt{\tau} J_\nu (k\tau) + C_2 \sqrt{\tau} Y_{\nu} (k\tau) \, ,
\end{equation}
where
\begin{equation}
\nu^2 = \frac14 + \mu = \frac{n^2}{4} \, .
\end{equation}
We have
\begin{equation}
d_k \sim C_1 \tau^{\frac12-\nu} + C_2 \tau^{\frac52-\nu}
\end{equation}
as $t\to+\infty\Leftrightarrow \tau\to 0^-$, whence
\begin{equation}
d_k \sim C_1 e^{(\nu-\frac12)t} + C_2 e^{(\nu-\frac52)t} \, .
\end{equation}
This leads to
\begin{equation}
c_k \sim C_1 e^{-\frac{n-1}{2}t+(\nu-\frac12)t} + C_2 e^{-\frac{n-1}{2}t+(\nu-\frac52)t} \, ,
\end{equation}
that is,
\begin{equation}
c_k \sim C_1 + C_2 e^{-2t} \, ,
\end{equation}
implying in particular that
\begin{equation}
|\dot{c}_k| \lesssim e^{-2t} \, .
\end{equation}
Comparing with \eqref{mainestimatedS}, we see that \eqref{mainestimate} is (almost) sharp in the case $a(t)=e^t$. For a Fourier component analysis along similar lines in the RNdS case see~\cite{pedroPhd}.
%
%
\section{The conformally invariant wave equation}\label{appendixB}
It is interesting to contrast the behavior of solutions of the wave equation and the conformally invariant wave equation, which can be easily expressed in terms of solutions of the wave equation in the Minkowski spacetime.

The conformally invariant wave equation in $n+1$ dimensions is (see, for instance, \cite{Wald})
\begin{equation}
\left(\Box_g - \frac{n-1}{4n} R_g \right) \phi = 0 \, ,
\end{equation}
where $R_g$ is the scalar curvature of the metric $g$. If $g$ is a FLRW metric with flat $n-$dimensional spatial sections, then it is conformally related with the Minkowski metric,
\begin{equation} \label{FLRWmetric}
g = a^2(t) \left[ -d\tau^2 +  \left(\left(dx^1\right)^2 + \ldots + \left(dx^n\right)^2 \right) \right] \, ,
\end{equation}
with
\begin{equation}
\tau = \int \frac{dt}{a(t)} \, ,
\end{equation}
and so any solution of the conformally invariant wave equation is of the form
\begin{equation}
\phi = a^{1-\frac{n+1}2} \psi \, ,
\end{equation}
where $\psi$ is a solution of the wave equation in Minkowski spacetime. Thus, we have
\begin{equation}
\dot{\phi} \sim a^{-\frac{n+1}2} \dot{a} \psi + a^{1-\frac{n+1}2} \frac{\partial \psi}{\partial \tau} \frac{d\tau}{dt} = a^{-\frac{n+1}2} \left(\dot{a} \psi + \frac{\partial \psi}{\partial \tau} \right) \, .
\end{equation}
Both $\psi$ and its time derivative are bounded for any topology of the flat spatial sections, and so
\begin{equation}
|\dot{\phi}| \lesssim a^{-\frac{n+1}2} \left(\dot{a} + 1 \right) \, .
\end{equation}
If $a(t)=t^p$, then, for $p \leq 1$, we have
\begin{equation}
|\dot{\phi}| \lesssim a^{-\frac{n+1}2} \, ,
\end{equation}
replicating what was seen for the Fourier modes of the wave equation. For $p>1$, however, we have
\begin{equation}
|\dot{\phi}| \lesssim a^{-\frac{n+1}2} t^{p-1} = a^{-\frac{n-1}2-\frac1{p}} \, ,
\end{equation}
quite different from the behavior of the wave equation. If $a(t)=e^t$, we have
\begin{equation}
|\dot{\phi}| \lesssim a^{-\frac{n-1}2} \, ,
\end{equation}
again quite different from the behavior of the wave equation. Note that for the metric \eqref{FLRWmetric} we have
\begin{equation}
R_g = \frac{2n\ddot{a}}{a} + \frac{n(n-1)\dot{a}^2}{a^2} \, ,
\end{equation}
and so
\begin{equation}
R_g = \frac{np\left[(n+1)p-2\right]}{t^2}
\end{equation}
for $a(t)=t^p$, and
\begin{equation}
R_g = n(n+1)
\end{equation}
for $a(t)=e^t$.
%
%

\end{document}